\documentclass[prb, aps, twocolumn, superscriptaddress,noshowpacs]{revtex4}
\usepackage[english]{babel} 
\usepackage[english]{layout}
\usepackage{amsmath,amsfonts,amssymb}
\usepackage{graphicx}
\usepackage{mathrsfs}
\usepackage{float}
\usepackage{color}
\usepackage{braket}
\usepackage{version}
\usepackage{array,multirow,makecell}
\usepackage[T1]{fontenc}

\begin{document}

\title{Design of a room-temperature topological exciton-polariton laser in a ZnO/TiO$_2$-photonic crystal slab}
\author{I. Septembre}
\affiliation{Universit\'e Clermont Auvergne, Clermont Auvergne INP, CNRS, Institut Pascal, F-63000 Clermont-Ferrand, France}
\author{C. Leblanc}
\affiliation{Universit\'e Clermont Auvergne, Clermont Auvergne INP, CNRS, Institut Pascal, F-63000 Clermont-Ferrand, France}
\author{L. Hermet}
\affiliation{Universit\'e Clermont Auvergne, Clermont Auvergne INP, CNRS, Institut Pascal, F-63000 Clermont-Ferrand, France}
\author{H. S. Nguyen}
\affiliation{Univ Lyon, Ecole Centrale de Lyon, INSA Lyon, Universit\'e  Claude Bernard Lyon 1, CPE Lyon, CNRS, INL, UMR5270, Ecully 69130, France}
\affiliation{Institut Universitaire de France (IUF), 75231 Paris, France}
\author{X. Letartre}
\affiliation{Univ Lyon, Ecole Centrale de Lyon, INSA Lyon, Universit\'e  Claude Bernard Lyon 1, CPE Lyon, CNRS, INL, UMR5270, Ecully 69130, France}
\author{D. D. Solnyshkov}
\affiliation{Universit\'e Clermont Auvergne, Clermont Auvergne INP, CNRS, Institut Pascal, F-63000 Clermont-Ferrand, France}
\affiliation{Institut Universitaire de France (IUF), 75231 Paris, France}
\author{G. Malpuech}
\affiliation{Universit\'e Clermont Auvergne, Clermont Auvergne INP, CNRS, Institut Pascal, F-63000 Clermont-Ferrand, France}

\begin{abstract}
We propose theoretically a scheme to get a room-temperature 2D topological exciton-polariton laser with propagating topological lasing modes.   
The structure uses guided modes in a photonic crystal slab. A ZnO layer provides strong excitonic resonances stable at room temperature. It is capped by a TiO$_2$ layer pierced by a triangular lattice. The exciton-polariton modes of the 3D structure are computed by solving numerically Maxwell's equations including the excitonic response. The designed triangular lattice shows a transverse electric gap. The triangular lattice is shown to be the limit of a staggered honeycomb lattice when one of the sub-lattices vanishes. Its topology can be characterized by symmetry indicators. The interface between two shifted triangular lattices supports two counter-propagating modes lying in the gap of the bulk modes. The interface states are analogous to quantum pseudospin Hall interface states. These modes show orthogonal polarizations. They can be selectively excited using polarized excitation and are well-protected from back-scattering. These modes can benefit from the exciton-polariton gain at room temperature because of their sufficiently large exciton fraction and favorable position in energy. The strong localization of these propagating modes makes them suitable to host topological lasing triggered by a non-resonant pump localized on the interface.
\end{abstract}

\maketitle

\section{Introduction}
Topology is one of the most active fields of research in modern physics. Appearing as a field in the XIX century, it started to grow faster at the end of the XXth century~\cite{thouless1982quantized,berry1984quantal}, first to explain solid-state phenomena~\cite{haldane1988model}, and then extended to topological photonics~\cite{haldane2008possible,raghu2008analogs,ozawa2019topological}. Topological singularities, such as Dirac points~\cite{novoselov2005two,young2012dirac,young2015dirac,armitage2018weyl}, Weyl points~\cite{yan2017topological,lv2015experimental,soluyanov2015type,riwar2016multi,septembre2022weyl}, or exceptional points~\cite{Voigt1902,Richter2019,Liao2021,krol2022annihilation}, carry a topological charge describing how the eigenstates evolve critically close to the singularity~\cite{Mera2021}. The large variety of singularities explains that there are many different topological classes and phases~\cite{PhysRevB.78.195125,fidkowski2010effects,chiu2016classification,kruthoff2017topological}.

Topological photonics has been initiated by Haldane and Raghu~\cite{haldane2008possible,raghu2008analogs} and Solja\v{c}i\'{c}'s group~\cite{wang2008reflection,wang2009observation}. They proposed to break time-reversal symmetry (TRS) in photonic crystal slabs (PCS) in order to mimic the quantum anomalous Hall effect (QAHE)~\cite{haldane1988model} realizing one-way edge modes. This requirement to break TRS first implied working with gyromagnetic materials typically at microwave frequencies. The extension of this regime toward optical frequencies and the key role played by the transverse electric-transverse magnetic (TE-TM) photonic spin-orbit coupling~\cite{kavokin2005optical} emerged by considering the properties of exciton-polaritons modes~\cite{nalitov2015polariton,klembt2018exciton}. The key advantage of this broken-TRS phase is that it allows realizing truly topologically protected one-way modes. The disadvantage is that applying magnetic fields remains inconvenient for future applications, such as integrated photonic circuits.
As a consequence, another class of topological phases, which can be generically labeled quantum pseudospin Hall effect, became extremely popular in photonics~\cite{ozawa2019topological}. In analogy with the quantum spin Hall effect~\cite{PhysRevLett.95.146802,Hasan2010}, each pseudospin component of a two-level system is characterized by a topological invariant, which is changing sign through an interface supporting a pseudospin current. These pseudospins can represent the valley degree of freedom in a staggered honeycomb lattice~\cite{xiao2007valley}, the angular momentum of ring resonators~\cite{hafezi2011robust}, and even light polarization~\cite{khanikaev2013photonic} in systems where the permittivity equals the permeability $\epsilon=\mu$ and where TE-TM splitting is suppressed, or even $p$ and $d$ orbitals in shrunken-expanded honeycomb lattices~\cite{wu2015scheme,barik2016two}. In all these cases, the two pseudospin components must be uncoupled, which is the case if some symmetries are preserved, for example, a crystalline symmetry in the quantum valley Hall effects and shrunken-expanded honeycomb lattices. These modes have interesting properties, such as the possibility to go through sharp corners (of 120 degrees, which preserves the valley), but these interface modes are \textit{a priori} not protected from random local fluctuations of the Hamiltonian (structural disorder), which are necessarily present in real structures. However, it turned out that in staggered and shrunken-expanded honeycomb lattices based on photonic crystal slabs the valley pseudospin is coupled to the circular polarisation degree of light~\cite{barik2018topological} providing extra protection against inter-valley scattering.

Recent works have also shown that a staggered honeycomb lattice in PCS is even not required to get valley-polarized interface states and that using interfaces between triangular lattices is sufficient~\cite{zhou2021chip,yang2021evolution,davis2022topologically}. In this case, it is not possible to define a pseudospin Chern number and the quantity, which is discontinuous through the interface is a so-called symmetry indicator~\cite{ono2018unified,luo2021multi,wen2022designing}. 
This type of lattice is very advantageous from a technological point of view, since it removes the necessity to create the smallest of the two types of holes of a staggered honeycomb lattice PCS. 

One of the most emblematic devices born of the topological photonic concept is the topological laser, where lasing occurs on a topological edge or interface states. It took some time for the community to propose this concept, probably because topological photonics was initially developed in a wavelength range for which gain is essentially absent. A topological laser was first proposed in June 2015 based on a 1D SSH chain of 0D exciton-polariton modes~\cite{solnyshkov2016kibble}, which is a type of system where lasing occurs quite naturally~\cite{Imamoglu1996,kavokin2017microcavities}. It was then proposed in February 2016 in purely photonic 1D lattices~\cite{pilozzi2016topological}, where the name "topological lasing" was introduced first. It has been experimentally realized soon after at low temperature in polaritonic system~\cite{st2017lasing,dusel2021room,harder2021coherent} using etched microcavities. Quite simultaneously (June 2016), 2D topological laser hosting propagating edge or interface modes have been proposed~\cite{harari2016topological} and realized either in broken TRS phases (QAHE)~\cite{bahari2017nonreciprocal} or in a quantum pseudospin Hall effect setting~\cite{bandres2018topological}. Since then, the field considerably expanded. One can cite the realization of electrical pumped 2D topological lasers first at low temperature using valley edge modes~\cite{zeng2020electrically}, and then at room temperature with ring resonator lattices~\cite{choi2021room}. In strongly coupled polaritonic systems, 1D room-temperature topological polariton laser~\cite{dusel2021room} showing high coherence~\cite{harder2021coherent} has been demonstrated in organic-based systems. 2D topological lasers based on coupled-cavity lattices under magnetic field were proposed~\cite{kartashov2019two}, but in a scheme typically limited to low temperatures. Quantum pseudospin Hall effect has been implemented at room temperature using transition metal dichalcogenide monolayers placed on PCSs~\cite{li2021experimentalZ2}, but lasing has not yet been demonstrated. Historically, the achievement of room temperature polariton lasing relies either on using large band gap semiconductors, ZnO and GaN, first in microcavities~\cite{zamfirescu2002zno,malpuech2002polariton,christopoulos2007room,Feng2013}, then using guided polariton modes~\cite{jamadi2018edge,Guillet2022,delphan2022polariton} which allows long propagation distances ($\sim100\,\mu$m).

In this work, we propose a feasible design of a 2D room-temperature topological polariton laser with propagative interface states. The waveguide structure is composed of a ZnMgO cladding, of a ZnO layer providing strong and stable excitonic resonances, and of a TiO$_2$ layer with a high refractive index. The latter is etched with a triangular lattice of circular holes. 
We solve numerically Maxwell's equations by finite element methods for a 3D structure periodic in the ($x,y$) plane and find the dispersion of 3D polaritonic modes, the excitonic resonance being taken into account in the permittivity. We find a gap in the TE-modes of width 50 meV and whose energy can be set up to 3.25~eV with an exciton fraction of gap edges modes around 0.2, which are favourable parameters to get room temperature polariton lasing in ZnO-based materials \cite{Feng2013,jamadi2018edge}. We then model a 3D structure hosting topological interface states by creating an interface between two triangular lattices with the same parameters, without resorting to a staggered honeycomb lattice. Numerical constraints do not allow finding directly the polaritonic modes in such a structure, so we compute the bare photonic modes and describe the coupling between those modes and the ZnO excitonic resonances through an effective Hamiltonian. We show that by exciting the interface with a well-defined circular polarisation, the propagation occurs only at the interface and in a unique direction with a  very good selectivity. We finally discuss the possiblity to trigger polariton lasing specifically at the interface states using a focused non-resonant optical pumping, because its overlap with the interface modes can be made considerably larger than with bulk states. The scheme we propose could be used for new developments in integrated photonics/polaritonics, that is on-chip integration of room-temperature topological polariton lasers.

\section{Topological interface in a triangular lattice}\label{sec1}
In this section, we demonstrate that an the interface between two triangular lattices (which have the same geometric parameters) can hosts propagative interface states in the gap of the bulk bands.

We begin by considering a \emph{honeycomb} lattice of circular holes. The TE band structure exhibits conical intersections (Dirac points) at the corners of the Brillouin zone (K and K')~\cite{wen2008two}. In a tight-binding description of such lattice, the two Dirac points are characterized by opposite winding $\pm 1$ of the sublattice pseudospin. The staggering of the lattice makes sites A and B different. It opens a gap at the two K and K' points also called valleys. The Berry curvature of bands is opposite in the two valleys, so a valley Chern number can be defined. It has opposite signs at K and K'. Making an interface between two lattices with opposite staggering and inverted valley Chern numbers realizes the so-called quantum valley Hall effect, where the direction of propagation of interface modes is associated with a given valley.

\begin{figure}[tbp]
    \centering
    \includegraphics[width=0.99\linewidth]{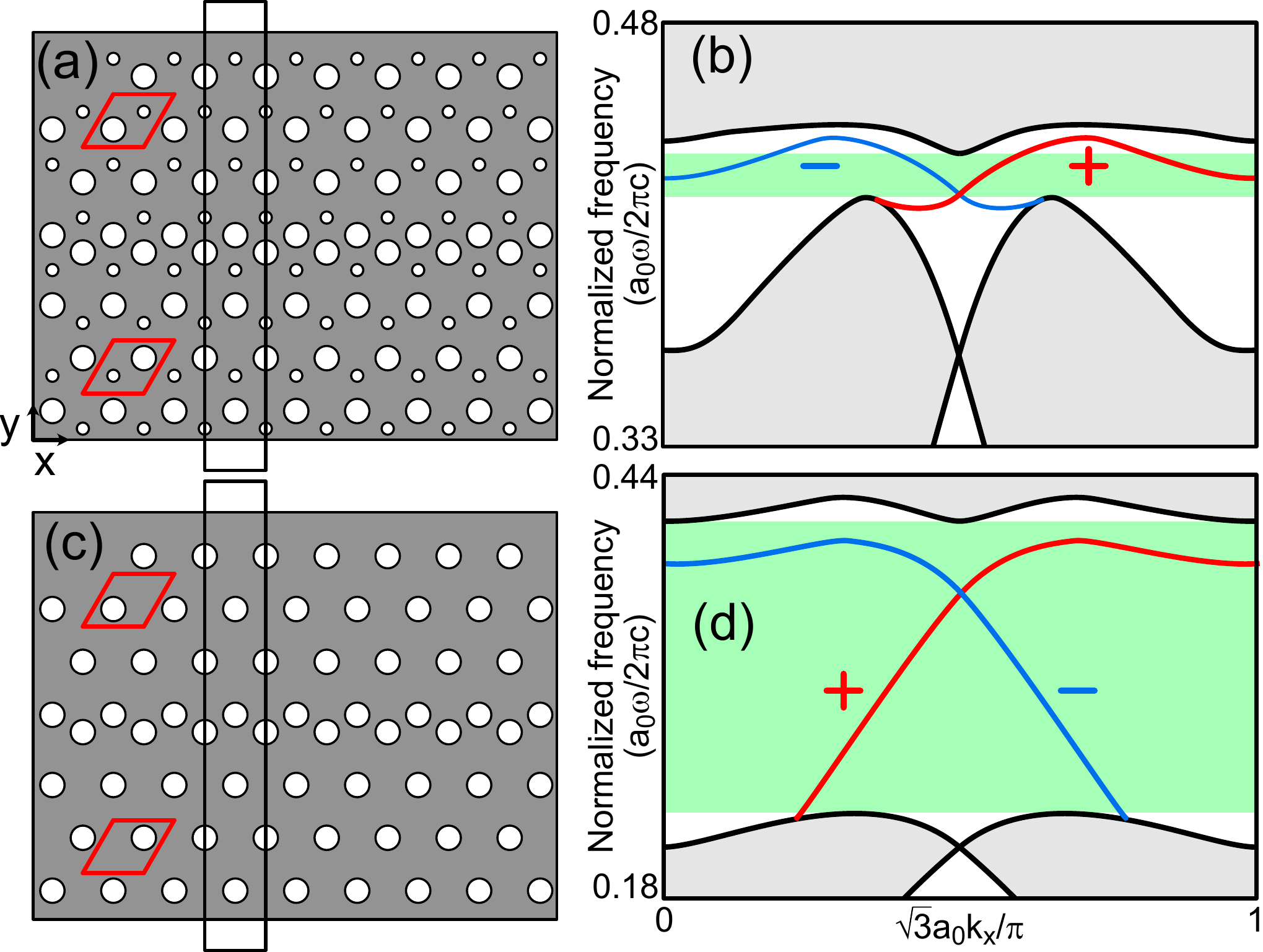}
    \caption{2D simulations demonstrating the presence of topological interface states even with a triangular lattice. (a,c) PC of circular air holes in a dielectric matrix and topological interface using a staggered honeycomb (a) and a triangular (b) lattice. The unit cell is emphasized in red, showing the difference between the upper and lower PCS. The triangular lattice is obtained by continuously reducing and ultimately removing the small hole of the honeycomb lattice. A ribbon of $a_0$ width along the $x$ direction and several periods in the $y$ direction, as used in the simulations to emulate an infinite system in the $x$ direction, is surrounded by black lines. (b,d) Band structures with interface states (blue line excited by a $\sigma_+$ polarized pump) in the gap (green area) of the structures with topological interfaces staggered honeycomb (b) and triangular (d) lattices. The bulk modes (grey areas) are delimited by black thick lines. Note that interface states exist in both cases whereas the gap is much larger in the triangular PC.}
    \label{fig1}
\end{figure}

Fig.~\ref{fig1}(a,b) shows the TiO$_2$ 2D photonic crystal (PC) we simulate (a) together with its dispersion (b) obtained by 2D simulations using COMSOL Multiphysics. In this section, we restrict our simulations to 2D structures for simplicity, because we focus on the effects of different kinds of patterning which are already visible in 2D. The software solves the Helmholtz equation:
\begin{equation}\label{Helm}
    \nabla \times (\nabla \times \mathbf{E}(\mathbf{r}))=k^2 \epsilon_\mathrm{r}(\mathbf{r})\mathbf{E}(\mathbf{r}),
\end{equation}
where $\mathbf{E}(\mathbf{r})$ is the electric field profile, $\epsilon_\mathrm{r}(\mathbf{r})$ the permittivity tensor, and $k$ the wavevector. It finds the spatial profiles and the energies of the eigenmodes.
The structure we simulated is schematically shown in Fig.~\ref{fig1}(a). It is a ribbon in the $y$ direction and an infinite structure in the $x$ direction using Floquet periodicity. The interface is a line in the $x$ direction.  We took a ribbon of 16 periods in the $y$ direction both for the PC above and below the interface. We can see that the upper PC has a staggering opposite to the one of the lower PC, as emphasized by the unit cells in red. The dispersion of the TE modes of the structure is plotted in Fig.\ref{fig1}(b) where the energy is calculated in reduced coordinates, the radii of the small and big holes being $R_-=0.17 a_0$ and $R_+=0.4 a_0$ respectively. 

We see that there is a broad energy range (green area) where no bulk states (grey areas surrounded by black lines) are present. This area is the band gap, where no bulk states are present. To have this gap close to the exciton energy in ZnO and GaN (around 3.2-3.4 eV) one needs to take $a_0\sim 100\,$nm corresponding to a small hole radius around $R_-\approx 15\,$nm, which is extremely challenging from a technological point of view.
Inside the gap, there are two interface states (red and blue lines) with non-zero group velocity. Because those states are in the gap of both the upper and lower PCs, they cannot scatter into the bulk states, so they propagate only at the interface with the direction of propagation associated with a valley.

The size of the gap is crucial in determining how well the interface states will be isolated from bulk states. It is determined by the radii of the two holes: if they are equal, the gap is null (there is a Dirac point), and the gap increases as they become more and more different~\cite{wen2008two,noh2020experimental}.
The limit $R_-=0$ is a triangular lattice of circular holes, as represented in Fig.~\ref{fig1}(c). It presents a particularity: a straightforward calculation gives zero Berry curvature, but there is no topological transition separating this limit from the non-trivial phase at $R_-\neq 0$. The symmetry indicators, calculated from the wavefunction at the high-symmetry points of the reciprocal space and allowing to determine the topology without integrating the Berry curvature over the whole Brillouin zone~\cite{Po2020}, also do not change~\cite{luo2021multi,zhou2021chip,wen2022designing}. We compute the dispersion in this case (with the same $R_+$) and plot it in Fig.~\ref{fig1}(d), and find that the interface states still exist in the gap, and the gap is much larger. This is in agreement with recent numerical and experimental studies~\cite{yang2021evolution,zhou2021chip,wen2022designing,davis2022topologically}.
Zero Berry curvature in the triangular lattice might be an artifact linked with the choice of the unit cell origin~\cite{bena2009remarks}. The staggered honeycomb and triangular lattices are topologically equivalent and going from one to another represents a continuous deformation.

In the following, we capitalize on this recently discovered behavior that facilitates the fabrication of structures and study a 3D polaritonic structure with two PCSs of simple triangular lattices of circular holes.

\section{Photonic crystal slab with exciton-polaritons in ZnO}
\begin{figure}[tbp]
    \centering
    \includegraphics[width=0.99\linewidth]{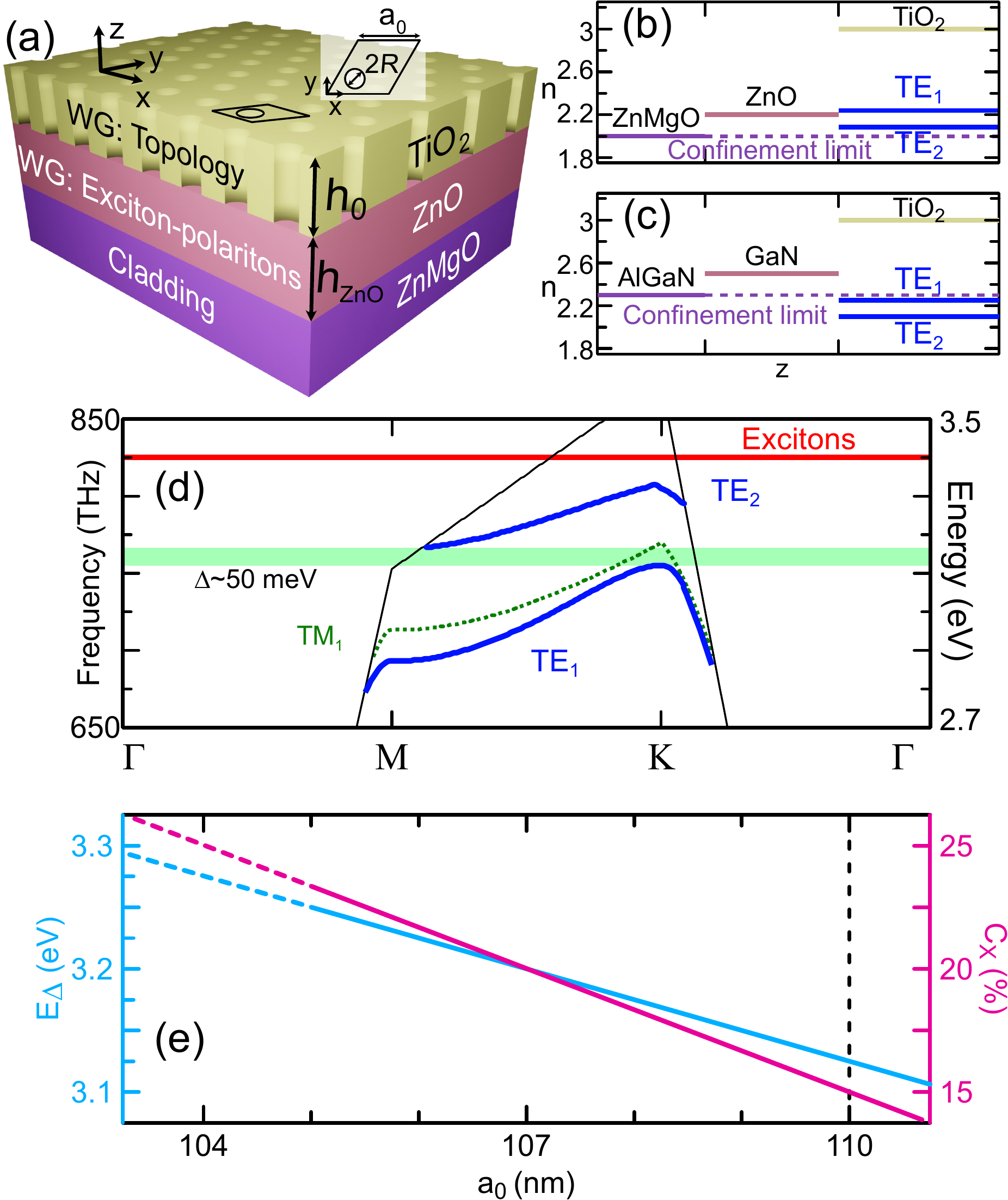}
    \caption{(a) Scheme of the PCS structure studied. A two-fold waveguide (TiO$_2$ in yellow and ZnO in pink) is separated from the substrate by a cladding layer (ZnMgO in purple). Only the TiO$_2$ layer is etched with a triangular lattice of circular holes. The thicknesses of the TiO$_2$ layer $h_0$ and ZnO layer $h_{\mathrm{ZnO}}$ are indicated. The 3D unit cell is shown in black and a 2D cut of it in the TiO$_2$ layer is shown in the inset with the geometric parameters of the lattice $R$ and $a_0$. (c,d) Refractive indices of the different layers in the $z$ direction for a structure using ZnO (b) and GaN (c). The effective indices of the TE$_{1,2}$ modes are indicated in blue. (d) Dispersion of the three modes below the exciton energy. There are two quasi-TE modes (blue solid lines) and one quasi-TM mode (green dashed line). The TE gap is emphasized by the light green area. Black lines represent the light cones, only modes below them are guided. Excitons energies are close and represented as a unique thick red line. (e) Energy of the center of the gap $E_\Delta$ and exciton fraction $C_X$ with respect to the period of the lattice $a_0$. The dispersions plotted in this work correspond to $a_0=110$~nm, represented as a vertical dashed line.}
    \label{fig2}
\end{figure}

The photonic crystal slab structure we consider is schematically depicted in Fig.~\ref{fig2}(a). It consists of a two-fold waveguide isolated from the substrate by a cladding layer (ZnMgO). The waveguide is made of patterned TiO$_2$ PCS (thickness $h_0$) and a bulk ZnO layer (thickness $h_{\mathrm{ZnO}}$). The PCS consists of a triangular lattice of circular holes. The lattice constant is $a_0$ and the diameter is $2R$. This part of the waveguide provides topology to the guided mode, whereas the ZnO part provides the strong coupling with an excitonic resonance, giving rise to exciton-polaritons.
The structure we consider directly comes out of the specifications explained hereafter. We want to build a structure capable of robust lasing behavior at room temperature so that it could be used in integrated photonics to pump photonic circuits. The lasing mechanism we want to use comes from the bosonic non-linearity of exciton-polaritons, giving rise to polariton lasing in the guided configuration~\cite{jamadi2018edge}. One, therefore, needs to fabricate a photonic crystal slab structure on a substrate, and not free-standing, to provide efficient heat dissipation. Moreover, the room temperature specification requires the use of wide-bandgap semiconductors, and the robustness restrains the choice essentially to GaN and ZnO. We have considered both and have finally chosen to focus on ZnO because of the following reasons.

Nowadays, ZnO can be grown on ZnMgO (itself grown on a ZnO or sapphire substrate) with a very good quality~\cite{herrfurth2021transient}. The ZnMgO layer serves as an optical cladding for the ZnO core, isolating it from the substrate, and as a buffer improving the growth quality. The best quality is obtained with $m$-plane ZnO~\cite{herrfurth2021transient}. However, the refractive indices of ZnO and ZnMgO are too close to each other, which prevents one from making a PCS by patterning directly the ZnO, because patterning makes its effective index smaller than that of ZnMgO, which suppresses the vertical confinement. Thus, we suggest using an extra layer with an effective index higher than that of ZnO for patterning. The TiO$_2$ is a particularly good candidate because of well-developed deposition and etching techniques. 

After deposition, the PCS is formed by etching only the TiO$_2$ layer. Close to the exciton resonance of ZnO ($E_X\approx3380\,$meV), the refractive index of TiO$_2$ is high ($n_{\mathrm{TiO}_2}\approx3$) and the losses are sufficiently low ($k_{\mathrm{TiO}_2}\approx10^{-4}$)~\cite{suppl}. Etching it can give a slab with an effective refractive of about 2.2, close to the one of ZnO at these energies~\cite{herrfurth2021transient}. 
We represent schematically the refractive indices of the different layers in Fig.~\ref{fig2}(b), showing that the etched TiO$_2$ layer provides light confinement for the two first TE modes TE$_{1,2}$.
A similar analysis is displayed in Fig.~\ref{fig2}(c) for an AlGaN/GaN/TiO$_2$ structure. It shows that TE modes are not confined in this structure, because of the excessive value of the AlGaN refractive index (for  $\sim 20\%$ of Al). A GaN-based structure would require a dielectric layer with a larger refractive index and small losses around the GaN exciton energy (around 3.5~eV).

In the simulations, we use frequency-dependent anisotropic permittivities for ZnO and ZnMgO layers~\cite{suppl} and isotropic for TiO$_2$. For ZnO, the $m$-plane growth brings in-plane anisotropy of both the background permittivity and the exciton resonances. Thus, in the $(x,y,z)$ basis, we have:
\begin{equation}\label{epsZnO}
    \epsilon_{\mathrm{Zn(Mg)O}}= \left( {\begin{array}{*{20}{c}}
\epsilon_{\mathrm{Zn(Mg)O}}^{||}(\omega)&0&0\\0&\epsilon_{\mathrm{Zn(Mg)O}}^{\perp}(\omega)&0\\0&0&\epsilon_{\mathrm{Zn(Mg)O}}^{\perp}(\omega)
\end{array}} \right).
\end{equation}
The ZnO exciton response is taken into account in the permittivity \cite{hopfield1958theory}:
\begin{equation}
    \epsilon_\mathrm{ZnO}(\omega)=\epsilon_\infty+\sum_{i=A,B,C}\frac{f_i}{\omega_i^2-\omega^2},
\end{equation}
where $A,~B,~C$ are the excitons of ZnO, $f_i$ their respective oscillator strengths, and $\omega_i$ their respective resonance frequencies. The non-radiative exciton lifetime can be added as an imaginary part.
The permittivity of Zn$_{1-x}$Mg$_x$O is taken from~\cite{teng2000refractive} for $x\approx0.2$, while the one of ZnO is extracted from~\cite{herrfurth2021transient}. For ZnO, the exciton resonances are located approximately at $3375$~meV (A), $3380$~meV (B), and $3410$~meV (C) at room temperature, while the oscillator strengths are approximately $f_A=150000$~meV$^2$, $f_B=250000$~meV$^2$ and $f_C=f_A+f_B$.

\begin{figure}[H]
    \centering
    \includegraphics[width=0.99\linewidth]{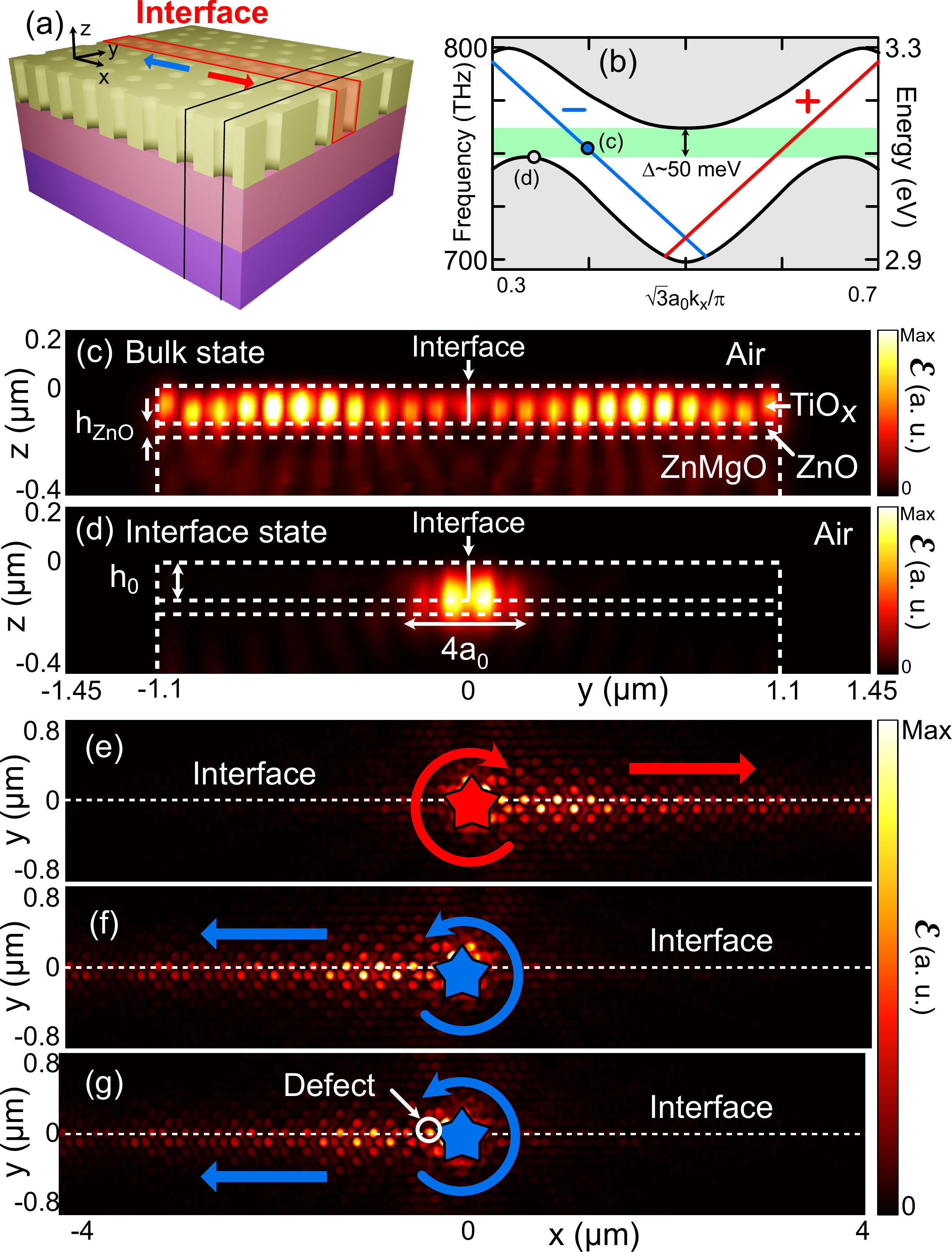}
    \caption{(a) Sketch of the 3D PCS structure hosting topological interface states. The slabs are the same as previously (TiO$_2$ in yellow, ZnO in pink, and ZnMgO in purple), but the top layer now contains an interface between two PCS of triangular lattices of circular holes (emphasized in red) where interface states can propagate in the left (blue) or right (red) direction. The black lines mark the 3D ribbon used in the simulations. (b) Polaritonic band structure calculated for a finite ribbon in the $y$ direction and infinite structure in the $x$ direction, including the interface. Note the interface states that can propagate in the gap formed by the bulk modes. An excitation with circular right (left) polarization leads to propagation in the right (left) direction (red/blue lines). (c,d) Electromagnetic energy magnitude profiles of the bulk state (c) and interface state (d) calculated from COMSOL. The corresponding states of these profiles in the dispersion (b) are indicated by the grey (blue) point for the bulk (interface) state. (e,f) Profiles of the interface states calculated by FDTD under a circular left-polarized excitation (e) and a circular right-polarized excitation (f) located below the interface. Note the propagation occurs mainly in the right/left direction (dashed white line). (g) Same as (f), but with a defect at the interface. Note the very good similarity between (g) and (f).}
    \label{fig3}
\end{figure}

Because we now consider a 3D structure, TE and TM modes are ill-defined, but we can still distinguish between quasi-TE and quasi-TM modes, the modes that are the 3D extensions of TE and TM modes, respectively. Numerically, COMSOL finds all photonic modes (quasi-TE and quasi-TM), and we only keep the quasi-TE modes by comparing the values of the electric and magnetic fields along the $z$ direction. For TE modes, $E_z=0$ and for TM modes, $H_z=0$. So, the 3D modes with a large ratio $|E_z|/|H_z|$ are the quasi-TM modes, whereas those where the ratio is small are the quasi-TE modes (the ones that we keep).

The dispersion of the quasi-TE polariton modes is plotted in Fig.~\ref{fig2}(d) for the path $\Gamma-M-K-\Gamma$ in the reciprocal space, which follows high-symmetry points of the lattice. Despite the broken symmetry along the $z$-direction and the presence of excitons, the dispersion is quite typical for this kind of PCSs, although flattened. Because we work with guided modes, we are interested in the modes that lie below the light cone. Moreover, we notice a gap between the two quasi-TE bands. This gap is centered around $3125\,$meV and of amplitude $\Delta\approx 50\,$meV. 
We calculated the exciton fraction $f_X$ of the modes near the gap edges by comparing the group velocity with excitons $v_X=\partial E_{\mathrm{ep}}/\partial k$ and without them $v_0=\partial E_{\mathrm{p}}/\partial k$, where $E_{\mathrm{ep}}$ is the energy of the lower polariton branch. In the end, we find $f_X=11\pm 2\,\%$ for the interface states, which is sufficiently large to observe polariton lasing~\cite{Feng2013,jamadi2018edge}.

The size of the gap reaches such value for the set of parameters  $h_{\mathrm{ZnO}}=50\,$nm, $h_{0}=130\,$nm, $a_0=110\,$nm and $2R=a_0/2=55\,$nm, according to our optimization study. Those dimensions, although challenging to obtain, must be achievable in state-of-the-art realizations, especially because we use a circular geometry for holes. The main challenge is to etch holes of such small diameter with a depth of more than 100\,nm. It could happen that the holes are not completely etched, meaning that they have the right diameter, but they do not reach the  ZnO layer. We performed additional simulations proving that a deviation of the order of a few nm of the depth of etching does not affect the results we present below. 

However, we noticed that under-etching (the situation that is more probable experimentally) is less deleterious than over-etching (which is anyway less probable experimentally). 
In the following, we consider that the TiO$_2$ slab is completely etched, and the ZnO layer remains intact.

There are no propagative quasi-TE states at the energies lying inside the gap. However, we show that there is still a quasi-TM mode that is inside the quasi-TE gap (in Fig.~\ref{fig2}(d), the dashed green line is the quasi-TM mode which is inside the green area, the quasi-TE gap). It has a very small overlap with the quasi-TE modes, so it will be disregarded as in the other works~\cite{barik2016two,barik2018topological,noh2020experimental}. A complete band gap for both quasi-TE and quasi-TM modes requires a very strong anisotropy of the refractive indices between in-plane and out-of-plane components~\cite{wen2008two}, and the anisotropy that we have in ZnO, despite being not negligible, is still too weak. However, we note that due to the exciton resonance, the second TM mode (the mode TM$_2$) is not present below the exciton resonance. This means that there is an effective TM gap starting from the TM$_1$ mode up to the next mode, which is above the exciton. This is an interesting feature, that we think may be used to create a PCS with a full quasi-TE and quasi-TM band gap.

Fig.~\ref{fig2}(e) shows the energy of the center of gap $E_\Delta$ with respect to the lattice constant $a_0$ for the same structure. The size of the hole is varied accordingly, in order to keep the filling factor of the TiO$_2$ layer constant. The closer to the exciton we are, the higher the energy and the exciton fraction. When the energies are too high ($E_\Delta>3.25\,$eV),  the modes cross the light cone, the confinement is lost and they penetrate into the cladding (the dashed lines continuing the solid lines). One can engineer the position of the center of the gap and the exciton fraction to get a polariton laser, as we will discuss later. The vertical dashed line indicates a period of $a_0=110\,$nm, which corresponds to the situation considered in Fig.~\ref{fig2}(d). This leads to an energy of the middle of the gap of $E_\Delta \approx 3125\,$meV and an exciton fraction of approximately $C_X\approx 15\%$.


\section{Interface states in the 3D PCS structure}
We now consider the structure discussed before, but the top layer is now composed of two triangular lattices shifted with respect to each other forming an interface. A scheme of the top layer is shown in Fig.~\ref{fig1}(c) and the full 3D structure is represented in Fig.~\ref{fig3}(a). Here, the symmetry in the $z$ direction is broken and one needs to check if the results of the 2D case (Fig.~\ref{fig1}) are still valid. The structure we consider is a triangular lattice of circular holes, infinite in the $x$ direction and 24 period-large in the $y$ direction. The upper half of the PCS is translated in the $y$ direction by $\delta y=-a_0\sqrt{3}/6$, which creates an interface between two triangular lattices. We use the same parameters as before, that is $a_0=110\,$nm and $2R=55\,$nm.

The knowledge of the permittivity of each material should be sufficient to find the dispersion of the structure, as we did for Fig~\ref{fig2}(d). However, the strong variation of the permittivity close to the exciton resonance prevents COMSOL from finding the eigenstates properly~\footnote{Close to the exciton resonance, the refractive index of ZnO varies rapidly with the energy and acquires large values, so the spatial mesh has to be extremely fine. Moreover, the structure containing an interface is approximately 20 times larger than a unit cell, which explains that the simulation time needed is larger for a structure with an interface than without).}. To circumvent this problem, we look for the dispersion of purely photonic modes $E_p(k)$ (neglecting the exciton resonance), and we post-process them to include properly the coupling to an effective excitonic resonance of energy $E_X=3375\,$meV. For that purpose, we use the matrix describing the strong coupling of excitons and photons~\cite{hopfield1958theory,kavokin2003cavity,haug2009quantum,kavokin2017microcavities}:
\begin{equation}\label{strongcoupling}
    M_{SC}= \left( \begin{array}{*{20}{c}}
E_X & \rho \hbar \Omega_R \\ \rho \hbar \Omega_R & E_p(k)
\end{array} \right),
\end{equation}
where $2\hbar \Omega_R=125\,$meV is the estimated Rabi splitting for the thicknesses of ZnO that we deal with~\cite{chen2009large,mihailovic2009optical,trichet2011one}, $E_X=3380\,$meV is the energy of the exciton and $\rho$ is the fraction of the mode confined in the ZnO layer. It is really important to take it into account because the waveguide we consider contains two layers, and thus an important part of the mode is not confined in ZnO, but rather in the PCS, which does not contain any exciton. We simulate this structure in COMSOL and find the dispersion $E_p(k)$ of photonic modes and their spatial distribution to extract $\rho$. The lower energy band (mode TE$_1$) is less confined in ZnO ($\rho\sim 10\%$) than the upper energy band (mode TE$_2$) for which the confinement in ZnO is approximately $\rho \sim 20\%$, which is also the case for the interface modes. We neglect the wave vector dependence of the exciton energy.

As a next step, we diagonalize the matrix~\eqref{strongcoupling} and find the polaritonic dispersion: 
\begin{equation}
    E_{\mathrm{LP}}=\frac{E_X+E_p(k)}{2}-\sqrt{(\rho\hbar \Omega_R)^2+\left(\frac{E_X-E_p(k)}{2}\right)^2}
\end{equation}
The corresponding band structure is plotted in Fig.~\ref{fig3}(c). It is consistent with the band structure found in Fig.~\ref{fig2}(d), but the structure is here infinite only in the $x$ direction. We still find a band gap for the bulk states at the same energy and of the same amplitude $\Delta \approx 50\,$meV, but now there are two modes in the gap of the bulk states, which are localized at the interface between the two triangular lattices. The two states are counter-propagating, one going in the $+x$ direction and the other going in the $-x$ direction, as expected. The group velocity of the interface states is $v_g=25\,\mu$m/ps, which is coherent with the existing literature~\cite{Ciers2020}.

In Fig.~\ref{fig3}(c,d), we show the spatial profile (obtained from the electromagnetic energy $|\mathcal{E}|$) in the ($y,z$) plane of the mode corresponding to the gray and blue points in Fig.~\ref{fig3}(b), respectively. We can see that the state corresponding to the gray point is not localized at the interface, but rather spread in the bulk of the PCS, while the state corresponding to the blue point is strongly localized at the interface, with a very narrow profile of only a few periods in the $y$ direction. We conclude that the blue line indeed corresponds to interface states, while the bulk states are in the grey regions, as expected.

So far, the interface modes are just two states of opposite wavevectors propagating in opposite directions. One could think that they could elastically scatter (by disorder) from one to another, which would give rise to Anderson-localized states. However, the interface state propagating to the right ($+$ states on  Fig.~\ref{fig3}(b)) is circularly polarized $\sigma^+$ on one side of the interface and $\sigma^-$ on the other side. The counter-propagating mode ($-$ states on  Fig.~\ref{fig3}(b)) shows the opposite polarisation pattern. These features can be found both from a tight-binding description of a staggered honeycomb lattice with TE-TM splitting and by examining the electric field pattern of modes numerically computed by COMSOL. These two counter-propagating states are orthogonal from a polarization point of view which prevents scattering from one to another by elastic scattering on the structural disorder. This protection from backscattering occurs for the vectorial electromagnetic field (photons) but does not \textit{a priori} hold for the electronic quantum valley Hall effect.

Next, we further illustrate this crucial property by simulating the propagation of wave packets on the interface.
This can be numerically simulated by using the Finite Difference Time Domain (FDTD) method. Time-dependent simulations are possible in COMSOL, but we resort to Lumerical, which is much more efficient for this kind of task. We used it to solve time-dependent Maxwell equations (including excitonic contributions in the permittivity given by Eq.~\eqref{epsZnO}) in this structure and probe the existence of interface states. By choosing the excitation position and polarization, it is possible to excite unidirectional interface states at a topological interface between two PCSs~\cite{gong2020topological}. We use this method and pump the interface with a circular left or circular right polarized electric dipole with frequency $f\approx 755\,$THz, corresponding approximately to the center of the gap (see Fig.~\ref{fig3}(b)). We then observe the propagation of the interface state for a few picoseconds and a few micrometers, as expected by the dispersion. In Fig.~\ref{fig3}(e-f), we plot the norm of the electromagnetic energy.
The image is plotted in the middle of the TiO$_2$ slab, $\sim 100\,$fs after the beginning of the simulation. The excitation pulse duration is chosen to be sufficiently long ($\delta \tau \sim 1\,$ps) to be narrow in frequency ($\delta f\sim 1\,$THz). We can see from the image that the propagation is mainly at the interface and that the signal propagates to the right (left) of the injection point if the excitation is polarized circular right in Fig.~\ref{fig3}(e) (left in Fig.~\ref{fig3}(f)), as expected~\cite{gong2020topological}. Indeed, we pump below the interface for both images, so that the direction of propagation of the topological interface states is given only by the polarization of the excitation. Note that exciting above the interface leads to inverted results, meaning that exciting circularly right (left) implies propagation to the left (right), because of the preserved chiral symmetry.

We calculate the directional selectivity (left-to-right) ratio:
\begin{equation}
    f_{\mathrm{L/R}}=\frac{P_L-P_R}{P_L+P_R},
\end{equation}
where $P_{\mathrm{L,R}}$ is the magnitude of the Poynting vector far from the injection ($\sim 4\,\mu$m far) integrated over a narrow 2D zone of few periods in both $x$ and $y$ directions and normalized. In the end, we find $f_{\mathrm{L/R}}=-0.93\pm 0.04$ for an excitation with circular right polarization and $f_{\mathrm{L/R}}=0.93\pm 0.04$ for an excitation with circular left polarization, which confirms a very high selectivity.

Moreover, we perform an additional simulation to verify that defects at the interface do not prevent topological interface states to propagate. We reproduce the simulation of Fig.~\ref{fig3}(f), but we double the radius of one hole at the interface, on the path of the propagating topological interface state. The profile of the mode is shown in Fig.~\ref{fig3}(g), where we can see that despite the defect indicated with a white circle, the profile of the mode is very similar. We calculated in this case the directional selectivity ratio and find $f_{\mathrm{L/R}}=0.90\pm 0.03$, which is a bit lower than the one found without defect, but still very close to one, as expected. We estimate that the back-scattering on the defect is about $3\%$, which is very low. This can be attributed to the topological nature of the states. However, it is not completely zero, because scattering to the opposite polarization is still possible although minimized by the polarization properties of the modes.

\section{Topological polariton lasing}
By itself, the presence of interface states does not ensure that there can be lasing from them.
There is first a need for gain, which can be electron-hole gain in a standard laser or polaritonic gain in a polariton laser. 
Room temperature polariton lasing in ZnO cavities~\cite{Feng2013} and waveguides~\cite{jamadi2018edge} have already been reported. In these references, lasing was achieved around 3.2 to 3.25~eV with exciton fractions $C_\mathrm{X}$ of the order of 20\%. In the previous sections, we considered a structure showing a gap at approximately 3.1-3.15~eV. Fig.~\ref{fig2}(e) shows the energy of the center of the gap $E_\Delta$ versus the lattice period $a_0$ for the structure without interface. We find that the upper side of the gap remains below the light cone up to 3.25~eV for the energy of the gap center. We conclude that a topological gap around 3.2~eV is feasible ($a_0=107\,$nm), which allows keeping the interface modes below the light cone, but it is the maximum value that can be achieved. The corresponding interface state shows an exciton fraction around 0.2, slightly larger than modes having the same energy in bulk cavities, because the overlap between the electric field and excitons is a little bit better in guided geometry. 
\begin{figure}[tbp]
    \centering
    \includegraphics[width=0.99\linewidth]{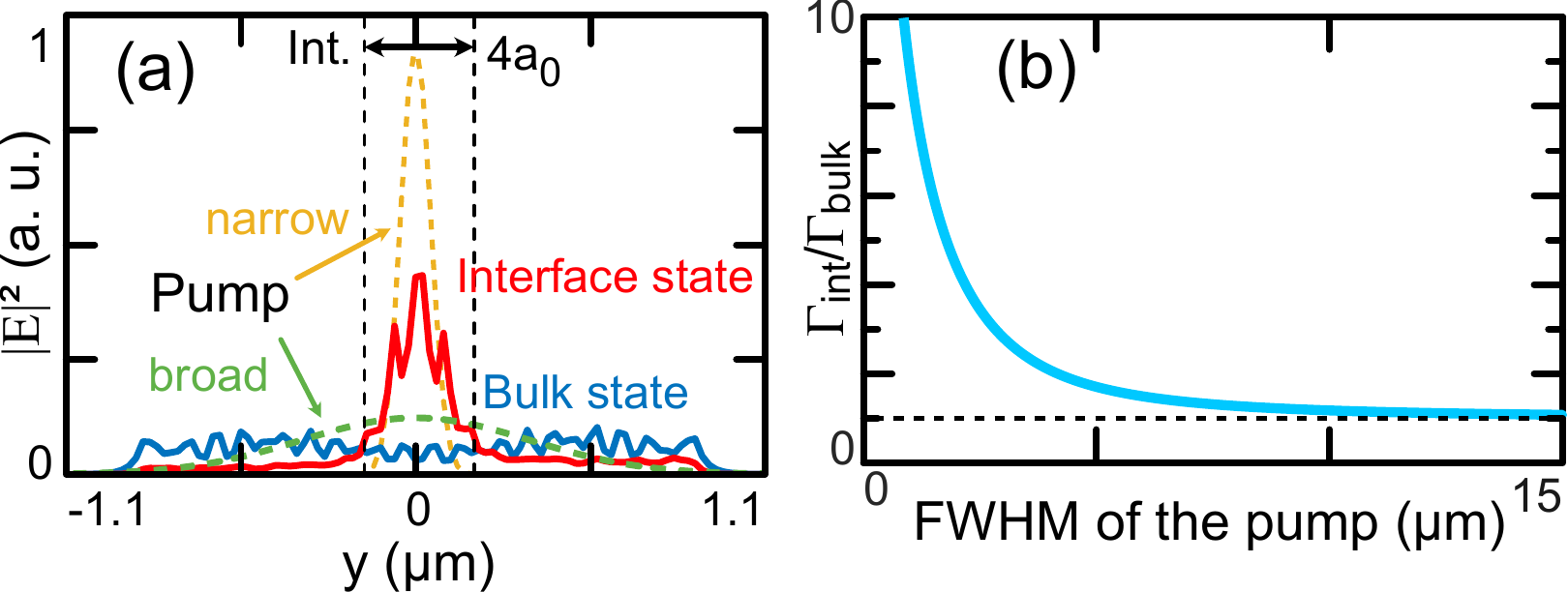}
    \caption{(a) Electric field profile (square amplitude) of the pump (dashed lines) and states (solid lines) in the $y$ direction. The norm of the electric field of the states is integrated along the $x$ and $z$ directions (and restricted to the ZnO) to give the profile. A broad pump has a good overlap with the interface states while the overlap is smaller for a narrow pump. (b) Ratio between the overlap of the pump profile with the interface $\Gamma_{\mathrm{int}}$ and bulk $\Gamma_{\mathrm{bulk}}$ states with respect to the size of the Gaussian pump, here its full width at half maximum (FWMH). A ratio of 1 is indicated by the dashed line.}
    \label{fig4}
\end{figure}

The second condition to get topological lasing is that mode competition should favor the topological interface states instead of bulk states. The most efficient approach is to focus the non-resonant pump laser on the topological state (on the interface), as proposed in~\cite{solnyshkov2016kibble} and done in~\cite{st2017lasing}. Fig.~\ref{fig4}(a) shows the spatial distribution along $y$ of the interface mode at 3.2~eV and of a bulk state. They exhibit a small overlap. The thin dashed lines represent two Gaussian excitations (narrow and broad). The ratio between the pump-to-interface $\Gamma_\mathrm{int}$ and the pump-to-bulk $\Gamma_\mathrm{pump}$ overlap versus the full width at half maximum (FWHM) of the Gaussian is shown in Fig.~\ref{fig4}(b). This ratio can be made arbitrarily large by considering a large sample, which increases the size of the bulk only. The qualitative conclusion is that a typical $\mu$m size pump laser excites the interface modes more than the bulk modes. For a pump smaller than $\sim 4 \,\mu$m, the overlap with the interface states is twice larger than with the bulk states. The overlap between the pump and the interface states is always better than with bulk states (the ratio $\Gamma_\mathrm{int}/\Gamma_\mathrm{pump}$ is always larger than 1, see the dashed line in Fig.~\ref{fig4}(b)), which favors lasing specifically on the interface states rather than on bulk states.

Another requirement is that lasing occurs on the interface states lying in the gap, rather than interface states outside the gap that are resonant with bulk modes. In practice, interface modes out of the gap are expected to suffer losses due to their coupling to the bulk modes by elastic scattering on the disorder, which is extremely significant in this frequency range so that the in-gap interface states are expected to be strongly favored.
On the other hand, both interface modes propagating in opposite directions are expected to be excited by the non-resonant pump, as was the case in previous papers reporting topological lasers based on quantum pseudo-spin Hall effect~\cite{bandres2018topological}. This can be overcome by using a circularly polarized non-resonant pump spatially shifted with respect to the interface, similarly to what is done in this work. Such pumping should favor one of the two interface modes leading to directional, provided the generated exciton reservoir does not lose completely its polarization.

\section{Conclusion}
To conclude, we propose a realistic design for a room-temperature 2D topological polariton laser. We model the full 3D structure of a photonic crystal slab including a ZnMgO cladding, an active ZnO layer, and a patterned TiO$_2$ layer. The full structure demonstrates a quasi-TE gap for bulk modes and topological interface states in this gap, whose energy and exciton fraction are optimal to get room temperature polariton lasing. We find that the topological interface states have an excellent one-way character upon appropriate excitation because they are protected from back-scattering by their polarization.

\begin{acknowledgements}
We thank N. Gippius, J. Zúñiga-Pérez, S. Bouchoule, E. Cambril, T. Guillet and J.-Y. Duboz for useful discussions.
This research was supported by the ANR Labex GaNext (ANR-11-LABX-0014), the ANR program "Investissements d'Avenir" through the IDEX-ISITE initiative 16-IDEX-0001 (CAP 20-25), the ANR project "NEWAVE" (ANR-21-CE24-0019) and the European Union's Horizon 2020 program, through a FET Open research and innovation action under the grant agreement No. 964770 (TopoLight).
\end{acknowledgements}

\bibliography{biblio} 
\end{document}